\documentclass[aps, prl, 10pt, twocolumn, superscriptaddress, noshowpacs, preprintnumbers, longbibliography, bibnotes,hyperref,floatfix]{revtex4-1}

\usepackage[colorlinks=true,breaklinks=true]{hyperref}
\hypersetup{allcolors=[rgb]{0.0 0.0 1.0},linkcolor=[rgb]{0.75 0.05 0.05}}
\usepackage[dvipsnames]{xcolor}
\usepackage{mathtools}
\usepackage{amsmath}
\usepackage{amsfonts}
\usepackage{amssymb}
\usepackage{bbold}
\usepackage{multirow}
\usepackage{bm}
\usepackage{hyperref}
\long\def\exclude#1{}
\usepackage{orcidlink}
\usepackage{slashed}

\DeclareRobustCommand{\okina}{%
  \raisebox{\dimexpr\fontcharht\font`A-\height}{%
    \scalebox{0.8}{`}%
  }%
}

\begin{document}

\title{Millicharged Particle Constraints from Asymptotic Giant Branch Stars
}

\author{Damiano F.\ G.\ Fiorillo \orcidlink{0000-0003-4927-9850}}
\email{damianofg@gmail.com}
\affiliation{Istituto Nazionale di Fisica Nucleare (INFN), Sezione di Napoli,
Complesso Universitario di Monte Sant’Angelo, Via Cintia, 80126 Napoli, Italy}
\affiliation{Gran Sasso Science Institute (GSSI), L’Aquila, Italy}

\author{Giuseppe Lucente
\orcidlink{0000-0003-1530-4851}}
\email{lucenteg@slac.stanford.edu}
\affiliation{SLAC National Accelerator Laboratory, 2575 Sand Hill Rd, Menlo Park, CA 94025}

\author{Jeremy Sakstein \orcidlink{0000-0002-9780-0922}} \email{sakstein@hawaii.edu}
\affiliation{Department of Physics \& Astronomy, University of Hawai\okina i, Watanabe Hall, 2505 Correa Road, Honolulu, HI, 96822, USA}

\author{Edoardo Vitagliano
\orcidlink{0000-0001-7847-1281}}
\email{edoardo.vitagliano@unipd.it}
\affiliation{Dipartimento di Fisica e Astronomia, Università degli Studi di Padova,
Via Marzolo 8, 35131 Padova, Italy}
\affiliation{Istituto Nazionale di Fisica Nucleare (INFN), Sezione di Padova,
Via Marzolo 8, 35131 Padova, Italy}

\begin{abstract}
We investigate the effect of millicharged particles (MCPs) with electric charge $qe\ll e$ and mass $m_\chi$ on the late-stage evolution phases of low-mass stars in globular clusters. We predict the $R_2$ parameter---the ratio of the number of stars in the asymptotic giant branch (AGB) phase to the number of stars in the horizontal branch (HB) phase---and compare it against globular cluster data. While the production of MCPs shortens both the HB and AGB lifetimes, a larger reduction in the AGB phase arises from the higher temperatures in the helium-burning shell.~We find the strongest bounds in the range $10\,\mathrm{keV}\lesssim m_\chi\lesssim 100\,\mathrm{keV}$, reaching charges as small as $q\simeq 5\times10^{-13}$ and surpassing existing constraints by up to two orders of magnitude.\end{abstract}


\maketitle

\textbf{\textit{Introduction.---}}A minimal extension of the Standard Model (SM) of particle physics is a sector featuring millicharged particles (MCPs), new fermions $\chi$
charged under a hidden $ U(1)_{\rm H}$ symmetry that couples them to a dark photon (DP) mixing kinetically with the SM photon~\cite{Galison:1983pa,Holdom:1986eq} (see also Refs.~\cite{Holdom:1985ag,Dienes:1996zr,Goodsell:2009xc,Albertus:2026fbe}),
\begin{equation}
    \mathcal{L}\supset-\frac{1}{4}F'_{\mu\nu}F'^{\mu\nu}-\frac{\epsilon}{2}F'^{\mu\nu}F_{\mu\nu}
    +\bar{\chi}(i\gamma^\mu\partial_\mu+g_\chi \gamma^\mu A'_\mu-m_\chi)\chi.
\end{equation}
The field redefinition $A_\mu'\rightarrow A_\mu'-\epsilon A_\mu$ brings the kinetic term into its canonical form, revealing that MCPs have a charge $g_\chi \epsilon=q e \ll e$,
\begin{equation}
\mathcal{L} \supset q e\,\overline{\chi} \gamma^\mu\chi\,A_\mu + \overline{\chi}(i\slashed{\partial}-m_\chi)\chi \, .
\end{equation}

Such simple dark sectors arise naturally in extensions of the Standard Model motivated by open questions in particle physics. For example, charge quantization appears to be supported by observations, yet its explanation requires physics beyond the Standard Model~\cite{Foot:1990mn}. If charge quantization is not a fundamental law of nature, then MCPs would arise naturally. In addition, MCPs provide a simple realization of dark sectors in which some or all of the dark matter (DM) carries a small electric charge, $qe\neq 0$, with $q\ll 1$~\cite{Essig:2011nj,Chu:2011be,Dvorkin:2019zdi,Iles:2024zka}. Discovering indications of MCPs could therefore shed light on the particle nature of dark matter.

\begin{figure*}
\includegraphics[width=1\columnwidth]{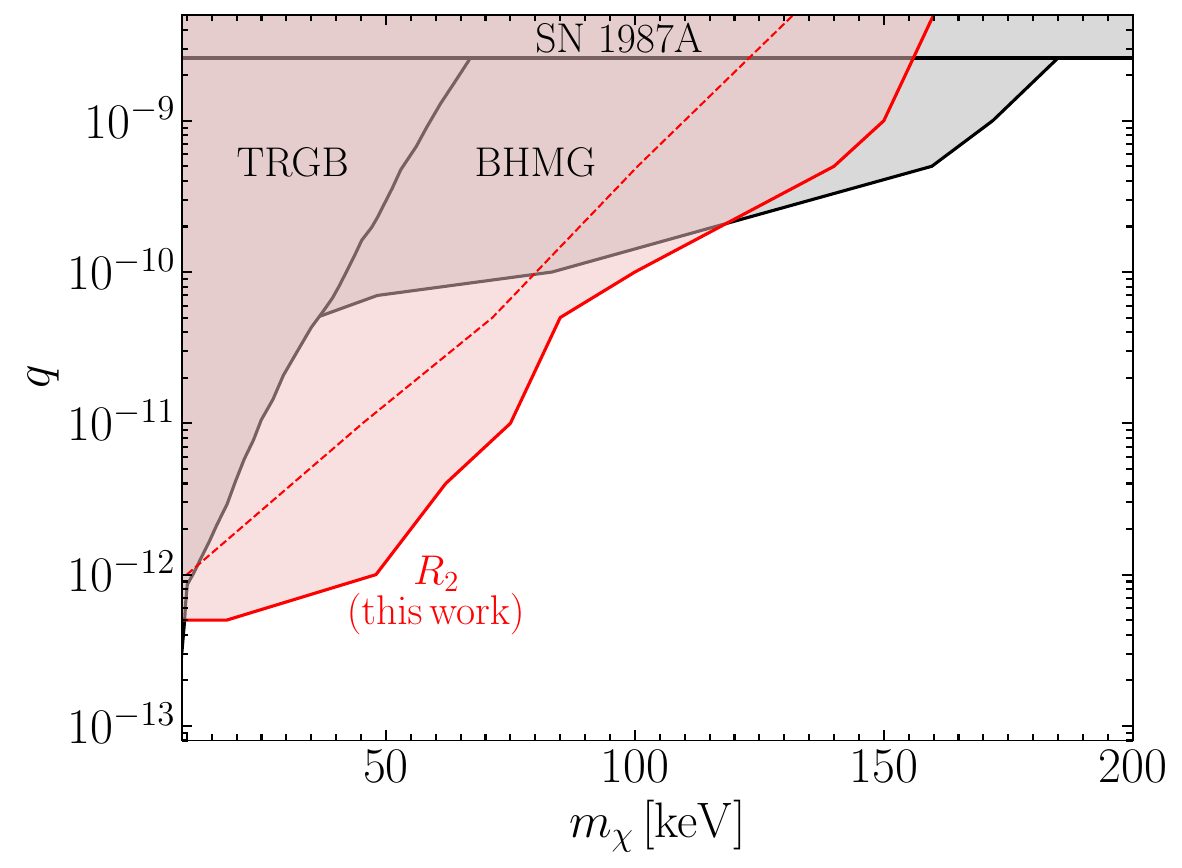}
\includegraphics[width=1\columnwidth]{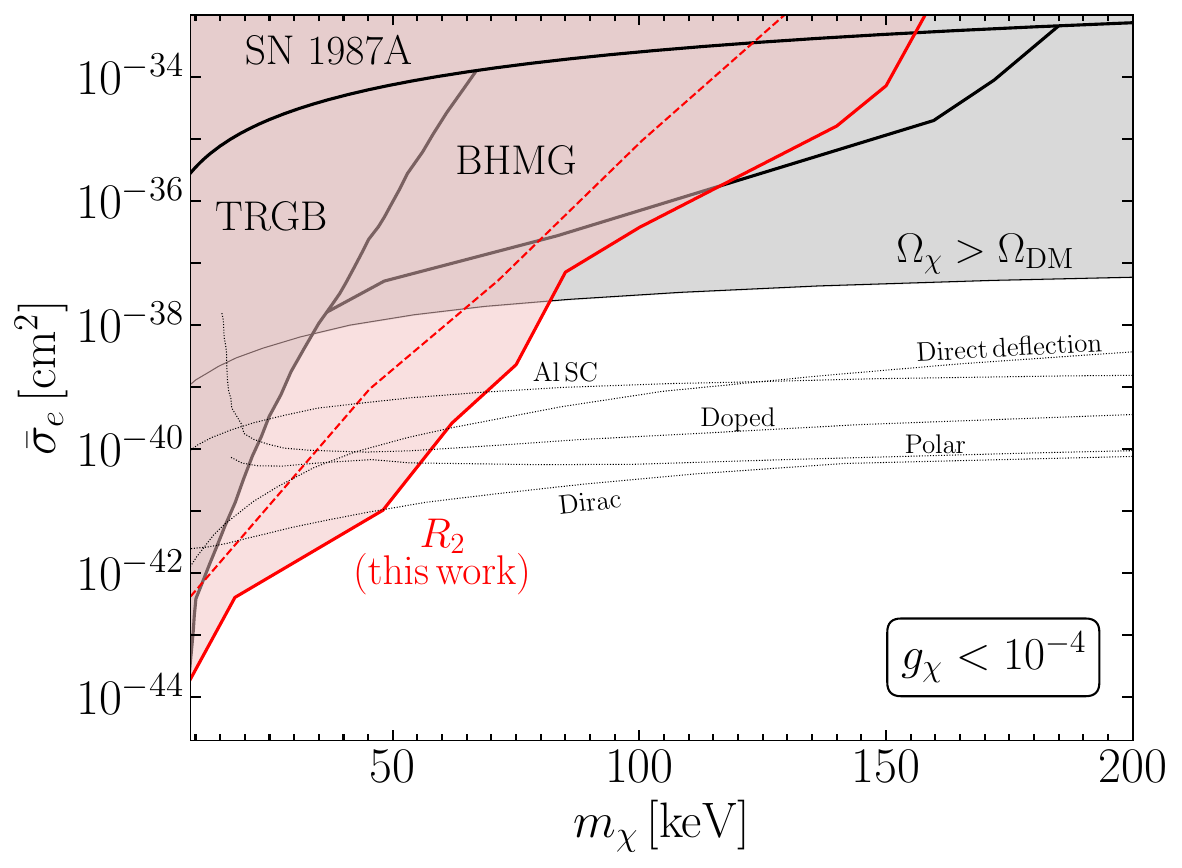}
    \caption{\textit{Left}: Constraints on the fractional charge $q$ as a function of the MCP mass; existing bounds are from the tip of the red giant branch~\cite{Fung:2023euv}, from the black hole mass gap~\cite{Fiorillo:2026vkd}, and from the observed neutrino cooling of SN~1987A~\cite{Fiorillo:2024upk}. The red dashed line shows an estimate for the effects of systematic stellar modeling uncertainties.
    \textit{Right}: Constraints on MCPs as a dark matter candidate; the DM abundance is obtained assuming the self-coupling to be $g_\chi\lesssim10^{-4}$~\cite{Iles:2024zka}. The sensitivity reaches of future experiments, including direct deflection~\cite{Berlin:2019uco}, polar materials~\cite{Knapen:2017ekk,Griffin:2018bjn}, Dirac materials~\cite{Hochberg:2017wce,Geilhufe:2019ndy}, doped semiconductors~\cite{Du:2022dxf}, and superconductors (Al SC)~\cite{Hochberg:2021pkt,Knapen:2021run}, are shown as dotted lines.
    }
    \label{fig:MCP_R2_bounds}
\end{figure*}

Consequently, MCPs have motivated extensive laboratory, cosmological, and astrophysical observations~\cite{Jones:1976xy,
Dobroliubov:1989mr,Mohapatra:1990vq,Davidson:1991si,Mitsui:1993ha,
Davidson:1993sj,Prinz:1998ua, Davidson:2000hf,Dubovsky:2003yn,Gies:2006ca,Gninenko:2006fi,
Badertscher:2006fm,Kim:2007zzs,
Jaeckel:2009dh,
Jaeckel:2010ni,Diamond:2013oda,
Vogel:2013raa,Moore:2014yba, Haas:2014dda,Vinyoles:2015khy,Chang:2018rso,Magill:2018tbb,Berlin:2018bsc,ArgoNeuT:2019ckq,Berlin:2019uco,Pospelov:2020ktu,Budker:2021quh,Carney:2021irt,ArguellesDelgado:2021lek,Berlin:2021kcm,milliQan:2021lne,Gan:2023jbs,Berlin:2024ewa,Iles:2024zka,CMS:2024eyx,Alcott:2025rxn,Berlin:2025btf,Berlin:2025hjs}. These include direct-detection experiments (see e.g.~\cite{Knapen:2017xzo,Oscura:2022vmi,SENSEI:2023zdf,DAMIC-M:2023gxo}) below $m_\chi\lesssim1\,\mathrm{MeV}$, and cosmological considerations, which imply that for these masses, regardless of the reheating temperature, an irreducible population of MCPs is produced in the early Universe, potentially overclosing the Universe for $q\gtrsim 10^{-11}$ if $g_\chi\lesssim 10^{-4}$~\cite{Iles:2024zka}.

Here we are interested in astrophysical probes of MCPs, which reach significantly smaller values of $q$. Different mass ranges are best probed by different sources. Heavy MCPs ($m_\chi\gtrsim 500\,\mathrm{keV}$) are most readily produced in the hot and dense core of supernovae (SNe), so stringent bounds come from the cooling of SN~1987A~\cite{Davidson:2000hf,Chang:2018rso,Fiorillo:2024upk} (see Ref.~\cite{Fiorillo:2023frv} for a discussion of the associated modeling uncertainties) and from low-energy SNe observations~\cite{Fiorillo:2024upk}, extending to masses of tens of MeV but only reaching down to $q_{\rm SN}\sim 10^{-9}$. At low masses ($m_\chi\lesssim 10\,\mathrm{keV}$), the tip of the red giant branch (TRGB) constrains charges as small as $q_{\rm TRGB}\sim 10^{-14}$~\cite{Fung:2023euv}.

In the intermediate mass range, $10\,\mathrm{keV}\lesssim m_\chi\lesssim 500\,\mathrm{keV}$, it was recently shown that late-stage stellar evolution is highly sensitive to MCP emission. The cooling rate of stellar cores in these stages due to MCP emission, calculated in Ref.~\cite{Fiorillo:2026shb}, depends sensitively on the temperature attained during helium burning.~In massive stars ($M\gtrsim 20\,{\rm M}_\odot$), it was found that MCP emission weakens (pulsational) pair-instability supernovae, predicting a shift of the lower edge of the upper black hole mass gap to higher masses, which led us to new and competitive bounds~\cite{Fiorillo:2026vkd}.

In this paper we show that the globular cluster $R_2$ parameter, the ratio of asymptotic giant branch (AGB) to horizontal branch (HB) stars, provides a sensitive probe of this intermediate MCP mass range. The sensitivity comes from the different helium-burning temperatures: MCP production can be Boltzmann-suppressed in HB cores, while remaining efficient in the hotter AGB helium-burning shells. Thus, MCP cooling changes the relative lifetimes of the two phases and reduces $R_2$. We use this effect to derive bounds in the range $10\,{\rm keV}\lesssim m_\chi\lesssim 100\,{\rm keV}$, improving over previous constraints by up to two orders of magnitude and reaching parameter space targeted by future searches for freeze-in millicharged dark matter, as shown in Fig.~\ref{fig:MCP_R2_bounds}.

\textbf{\textit{HB and AGB phases of stellar evolution.---}}The main phases of interest for the constraints we draw here are the HB and the AGB phases (see e.g. Ref.~\cite{Kippenhahn:2012qhp} for a textbook discussion). The HB phase follows the helium flash, which marks the end of the red giant branch. After helium ignites in the core, the star settles into a phase of stable core-helium burning. These stars occupy an approximately horizontal sequence in the Hertzsprung--Russell (HR) diagram:~their luminosities are primarily set by helium burning in the core, while their effective temperatures depend on the mass of the remaining hydrogen-rich envelope.~The core temperature remains of order $T\simeq 10^8\,\mathrm{K}$, as required for stable helium burning.

The HB phase ends once the central helium is exhausted, and the star enters the AGB phase, during which the core is composed of a carbon-oxygen core surrounded by a helium-burning shell with $T\simeq 3\times 10^8\,\mathrm{K}$. The $R_2$ parameter of a globular cluster is defined as the ratio of the number $N_{\rm AGB}$ of stars in the AGB phase to the number $N_{\rm HB}$ in the HB phase. Because the number of stars in a given evolutionary phase is proportional to the time spent in that phase, $R_2$ measures the ratio of the AGB to HB lifetimes $\tau_{\rm AGB}/\tau_{\rm HB}$.

\textbf{\textit{MCP emission from hot stellar cores.---}}MCPs can be thermally-produced inside stars and subsequently free-stream outwards, acting as a novel source of energy loss.~The rate of energy loss due to MCP emission from hot cores during the late stages of stellar evolution was recently derived in Ref.~\cite{Fiorillo:2026shb}. 
Analytic fits to the volumetric cooling rate as a function of temperature $T$, electron number density $n_e$, and the MCP mass $m_\chi$ and charge $q$ were provided;~we use these formulae in what follows.~To facilitate understanding the effects of MCPs on post-core helium burning, it is instructive to review the scaling of the loss rates via the various production channels with the interior stellar properties.

The dominant processes are semi-Compton emission $e^-+\gamma\to e^-+\overline\chi+\chi$ and pair production $e^-+e^+\to \overline\chi+\chi$. For Compton emission, the volumetric cooling rate scales as $Q_{\rm Com} \sim q^2 \alpha^3  n_e T^2 \mathrm{min}[1, T^2/m_e^2]e^{-m_\chi/T}$. 
The exponential factor is introduced as a rough description of the suppressed emission of heavy MCPs. 
In the non-relativistic limit relevant to the stars we study, the emissivity $Q_{\rm Com}$ grows as the fourth power of the temperature. This result is easy to obtain through dimensional analysis, noting that the Thomson cross section is proportional to $m_e^{-2}$.
For pair-production, the induced cooling rate is roughly independent of the density if degeneracy is neglected---larger electron chemical potentials imply a smaller fraction of positrons---so it scales as $Q_{\rm pair}\sim q^2 \alpha^2 T^5$ for $T\gg m_e$. While subdominant in the non-relativistic limit, this process is relevant in the AGB phase where the typical temperature $T\simeq 3\times 10^8\,\mathrm{K}\simeq 30\,\mathrm{keV}$ is only an order of magnitude below the electron mass.

\textbf{\textit{Impact of MCPs on stellar evolution.---}}~{MCPs shorten the nuclear-burning stages because the additional energy losses} must be compensated by faster burning rates. Due to the high temperature scaling of the MCP cooling rate (when $T\lesssim m_{\chi}$), we expect AGB stars to be more affected than HB stars, similar to the axion case~\cite{Dolan:2022kul}. This in turn implies that $R_2$ will be reduced since $R_2=N_{\rm AGB}/N_{\rm HB}\simeq\tau_{\rm AGB}/\tau_{\rm HB}$.

To study this quantitatively, we simulate the evolution of $0.82\,{\rm M}_\odot$ stars with metallicity $Z=10^{-3}$ and initial helium abundance $Y=0.254$ using MESA (version 12778) \cite{Paxton:2010ji,Paxton:2013pj,Paxton:2015jva,Paxton:2017eie,Paxton:2019lxx,MESA:2022zpy}, updated to include the MCP losses computed in Ref.~\cite{Fiorillo:2026shb}. The initial mass is representative of the progenitors of HB and AGB stars in old globular clusters, while the metallicity is typical of such systems. The adopted helium abundance is consistent with primordial helium inferred from low-metallicity H~II regions~\cite{Aver:2013wba}. The $R_2$ parameter is sensitive to $M$, $Y$, and $Z$ at the level of a few percent or smaller \cite{2016MNRAS.456.3866C,Dolan:2022kul}.

The largest sources of uncertainty in $R_2$ arise from the $^{12}{\rm C}(\alpha,\gamma)^{16}{\rm O}$ reaction rate, the treatment of convective mixing, and the timestep resolution~\cite{2016MNRAS.456.3866C}. The $^{12}{\rm C}(\alpha,\gamma)^{16}{\rm O}$ rate regulates the late stages of helium burning and hence the HB lifetime, while convective mixing determines the size and growth of the He-burning core. Timestep resolution affects the treatment of convective mixing and can induce {core breathing pulses}, further modifying the helium burning lifetime.  

In this work, we fix the nuclear and mixing prescriptions to standard choices, and assess the impact of numerical uncertainties by varying the resolution. Specifically, we adopt the default MESA nuclear rates (a combination of NACRE \cite{Angulo:1999zz} and REACLIB \cite{2016ApJ...830...55C}), and treat convection using the Henyey mixing-length scheme \cite{1965ApJ...142..841H} with $\alpha_{\rm MLT}=1.82$ and exponential overshooting with $f_0=f=0.001$ \cite{Dolan:2022kul}.

Our simulations run from the pre-main-sequence through the AGB. We define the zero-age horizontal branch (ZAHB) as the point where helium is being burned stably in the core, and the terminal age horizontal branch (TAHB) as the point where the central helium falls to $Y=10^{-4}$. The simulations are ended when the AGB mass loss reduces the envelope to 1\% of the total stellar mass.

We determine the $R_2$ parameter following Ref.~\cite{2016MNRAS.456.3866C,Dolan:2022kul}. First, for each set of MCP parameters we run 20 simulations, each with a different timestep on the HB in the range 1000 to 15000 years. For each simulation, we then construct a probability density function (PDF) for $\Delta\log L=\log L-\log L_{\rm HB}$, with $\log L_{\rm HB}$ the location of the HB peak in the luminosity PDF, as
\begin{equation}
    P(\Delta\log L)=\frac{1}{\tau}\sum_{i=1}^n\frac{\Delta t_i}{\sqrt{2\pi}\sigma}\exp\left(-\frac{(\Delta\log L-\Delta\log L_i)^2}{2\sigma^2}\right),
\end{equation}
where $n$ is the number of timesteps between the ZAHB and the point where $\Delta\log L=1.0$, $\Delta t_i$ is the time elapsed between models $i$ and $i+1$ (the timestep), $\tau=\sum_{i=1}^n\Delta t_i$ is the total time between $i=1$ and $i=n$, and $\sigma=0.02$ \cite{2016MNRAS.456.3866C}. Finally, we average over all 20 simulations to account for the uncertainty due to timestep resolution. We remove models exhibiting core breathing pulses, which are known to be sensitive to numerical and mixing prescriptions, and occur at a rate of $\le 1$ per parameter set.

\begin{figure}
    \centering
    \includegraphics[width=0.49\textwidth]{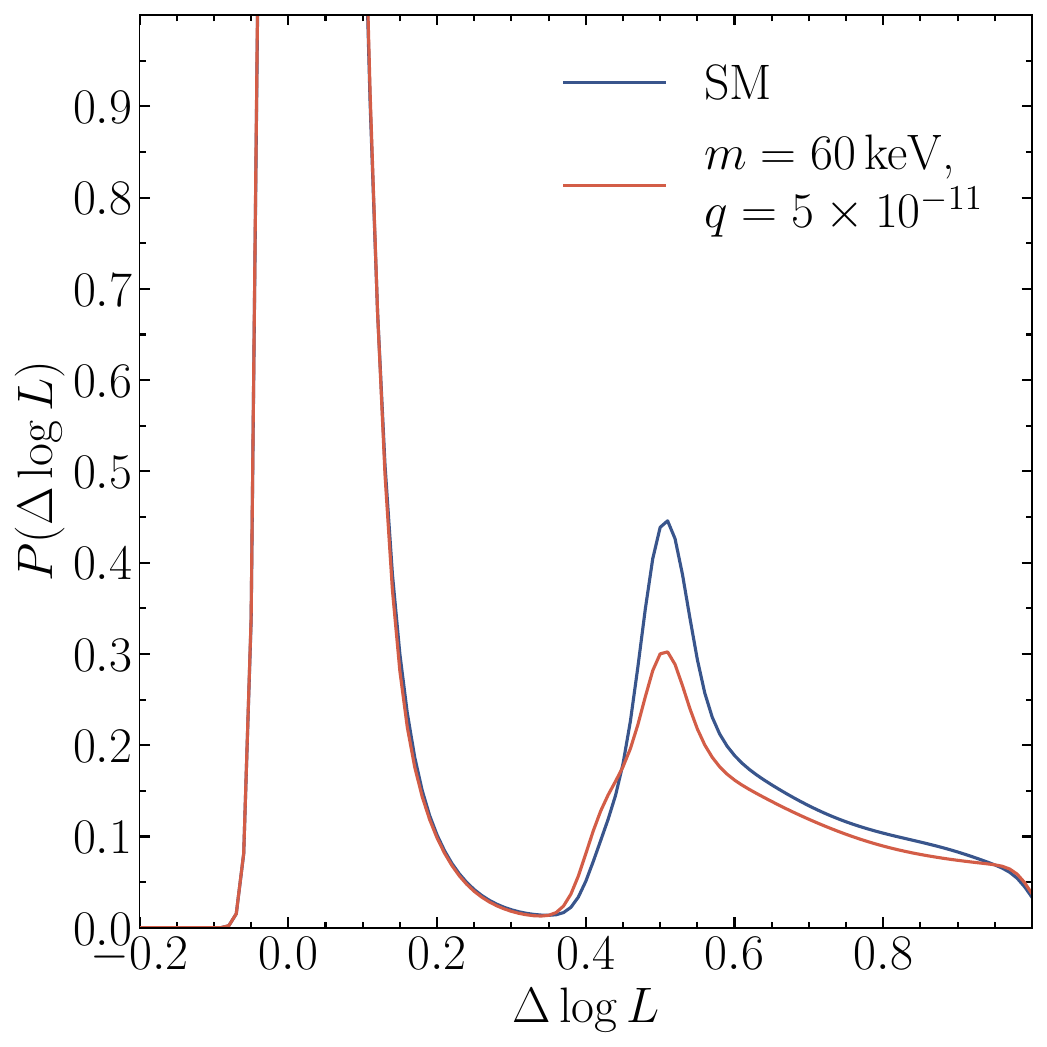}
    \caption{Example theoretical PDFs  for $\Delta\log L=\log L-\log L_{\rm HB}$ ($\log L_{\rm HB}$ is the location of the HB peak in the luminosity PDF) after averaging 20 simulations with varying timestep resolutions.}
    \label{fig:distributions}
\end{figure}

The impact of MCP emission is immediately visible in the PDF, which roughly measures the time span that the star spends in each luminosity interval. We show in Fig.~\ref{fig:distributions} the effect of MCP emission on the PDF for a representative choice of MCP charge and mass. The HB peak in the luminosity, due to the burning of the helium core, lies at $\Delta\log L\approx 0$. This is a very pronounced peak, due to the nearly constant luminosity across the relatively long HB phase. At higher luminosities, the star enters the AGB phase, exhibiting a smaller peak at $\Delta \log L\approx 0.5$. The $R_2$ parameter is given by $R_2=N_{\rm AGB}/N_{\rm HB}$, which we approximate as the ratio of the areas under the AGB and HB peaks, separated by the intervening minimum~\cite{2016MNRAS.456.3866C,Dolan:2022kul,Dolan:2023cjs}. The emission of MCPs with large masses leaves the HB peak mostly unaffected, but significantly alters the second peak, shortening the duration of the AGB phase, reducing the value of $R_2$. This differential sensitivity to the hotter AGB phase is the central observation of our work:~$R_2$ is especially powerful for probing heavier MCPs whose production is Boltzmann suppressed on the HB but remains efficient in the hotter helium-burning shell of AGB stars.

\begin{figure}
    \centering
    \includegraphics[width=0.49\textwidth]{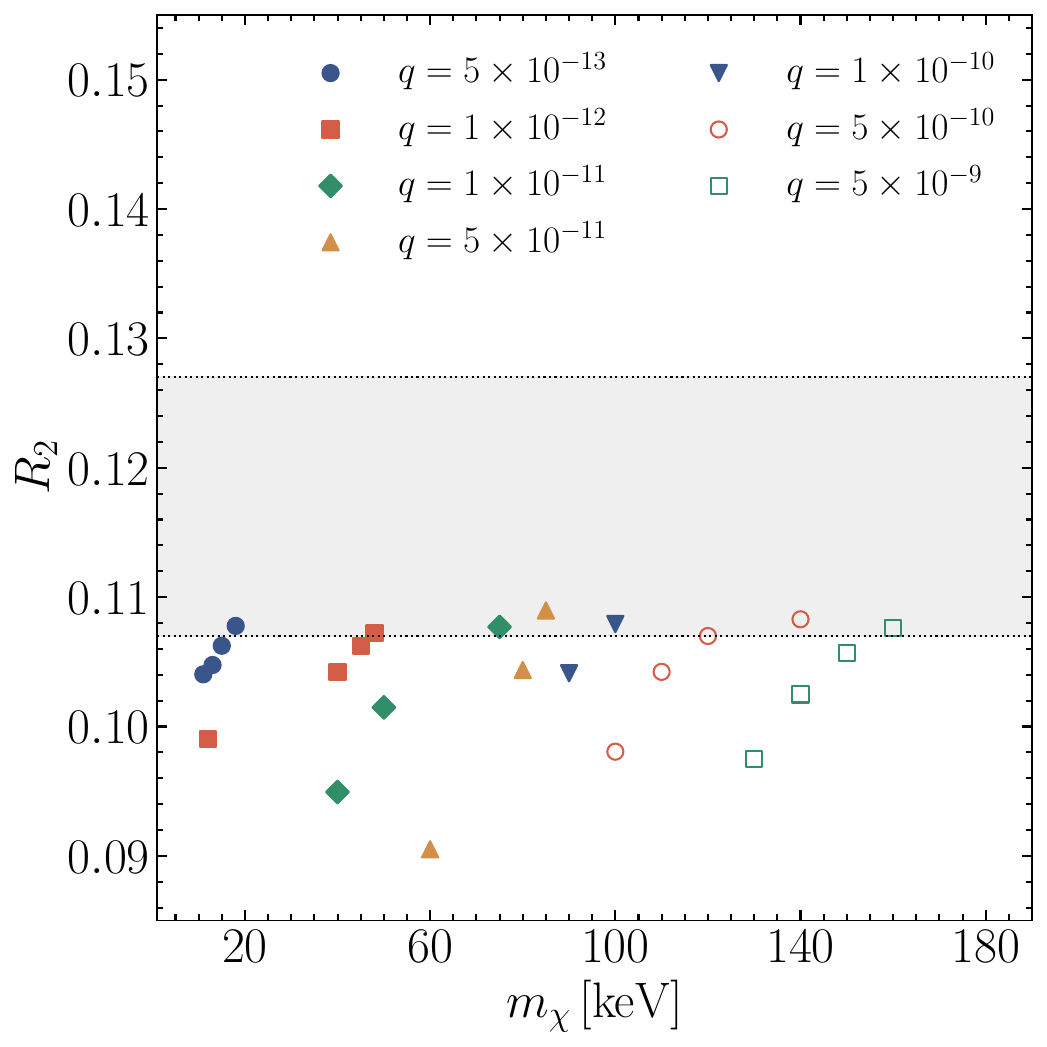}
    \caption{Predicted values of $R_2$ as a function of the MCP mass for different charges $q$ indicated in the figure. The gray region delimited by the dotted lines represents the 95\% C.L.~for the measured value.~MCP parameters that lie outside this band are excluded.
    } 
    \label{fig:R2_vs_MCP_parameters}
\end{figure}

\textbf{\textit{Results.---}}Our SM simulations yield $R_2=0.1084$, consistent at $1.72\sigma$ with the measured value \cite{2016MNRAS.456.3866C} $R_2=0.117\pm0.005$. Figure~\ref{fig:R2_vs_MCP_parameters} shows representative values of $R_2$ obtained when MCP losses were included with varying $m_\chi$ and $q$.~Evidently, for sufficiently large $q$ or small $m_\chi$ the predictions are incompatible with the measurement, so can be excluded. In the left panel of Fig.~\ref{fig:MCP_R2_bounds}, we show the constraints on $q$ (shaded red region) obtained by requiring that the
predicted \(R_2\) remains compatible with the observed value at \(95\%\) C.L.

\exclude{
\textbf{\textit{Results.---}}~Our SM simulations yield $R_2=0.1084$, consistent with the measured value \cite{2016MNRAS.456.3866C} $R_2=0.117\pm0.005$ at $1.72\sigma$. When MCP losses are included, both the HB and AGB lifetimes are reduced, with a larger reduction in the AGB phase due to the higher temperatures in the helium-burning shell. This is evident in Fig.~\ref{fig:distributions}, where the MCP peak is reduced relative to the SM.~Physically, MCP emission acts as an additional energy-loss channel, increasing the nuclear burning rate required for equilibrium and thereby shortening helium-burning lifetimes~\cite{Raffelt:1996wa}.~Since the AGB phase attains higher characteristic temperatures, the effect is enhanced in this phase.

For sufficiently strong couplings ($q$) or light masses $m_\chi$ (which reduce the Boltzmann suppression), the predicted value of $R_2$ is lowered such that it is no-longer consistent with the measured value.~This is shown in Fig.~\ref{fig:R2_vs_MCP_parameters}.~The bounds in Fig.~\ref{fig:MCP_R2_bounds} were found by demanding that MCP losses be compatible with the measured value at $2\sigma$.~}

While a full exploration of systematics is beyond the scope of this work, we estimate their potential impact by considering an optimistic recalibration in which the SM prediction for $R_2$ is shifted to the central measured value, and the same shift is applied to the MCP predictions. This assumes that the dominant stellar physics uncertainty acts as an approximately additive offset in $R_2$, with negligible covariance between the stellar physics and MCP cooling.~The red dashed line in Fig.~\ref{fig:MCP_R2_bounds} shows this recalibrated benchmark.~It corresponds to a weakening of the fiducial exclusion, but still excludes previously unconstrained regions of parameter space.

The right panel of Fig.~\ref{fig:MCP_R2_bounds} summarizes the impact of our constraints on MCPs as a dark-matter candidate in terms of the MCP--electron scattering cross section,
\begin{equation}
\bar{\sigma}_e=\frac{16\pi\,\alpha^2 q^2\,\mu_{\chi e}^2}{(\alpha m_e)^4}\,,
\end{equation}
where $\mu_{\chi e}$ is the MCP--electron reduced mass. Here the dark matter abundance is obtained assuming a sufficiently small self-coupling, $g_\chi\lesssim10^{-4}$. In this scenario, fractional charges $q\gtrsim10^{-11}$ are ruled out since the freeze-in MCP relic abundance would exceed the observed dark-matter density~\cite{Iles:2024zka}. Our $R_2$ bound probes below this cosmological ceiling and excludes MCPs as the dominant freeze-in dark matter component for ${m_\chi \lesssim 80\,{\rm keV}}$. This overlaps with parameter space targeted by next-generation direct-detection experiments, whose projected sensitivities are shown by the dotted curves~\cite{Berlin:2019uco,Knapen:2017ekk,Griffin:2018bjn,Hochberg:2017wce,Geilhufe:2019ndy,Du:2022dxf,Hochberg:2021pkt,Knapen:2021run}.

\textbf{\textit{Conclusions.---}}Millicharged particles (MCPs) arise in natural and motivated extensions of the Standard Model of particle physics. In this work, we have identified the $R_2$ parameter---the ratio of the number of stars in the Asymptotic Giant Branch (AGB) and Horizontal Branch (HB) phases---as a powerful probe of heavy MCPs ($10\,\textrm{keV}\lesssim m_\chi\lesssim 100\,\textrm{keV}$).~We obtained constraints in a regime where traditional stellar bounds are weakened by Boltzmann suppression.~The bounds extend to values as small as $q\lesssim 5\times 10^{-13}$, surpassing existing constraints by up to two orders of magnitude. These constraints have important implications for millicharged dark matter:~MCPs produced through freeze-in in the early Universe cannot constitute all of the dark matter for $m_\chi\lesssim 80\,\mathrm{keV}$, excluding part of the parameter space identified as a key target for future dark matter searches~\cite{Essig:2022dfa}.

Looking ahead, a systematic exploration of the stellar physics uncertainties entering $R_2$---including convective mixing and nuclear reaction rates---will be important for refining these bounds.

\textbf{\textit{Acknowledgments---}}This article is based on work from COST Action COSMIC WISPers
(CA21106), supported by COST (European Cooperation
in Science and Technology).
D.F.G.F. acknowledges support by the TAsP (Theoretical Astroparticle Physics) project, and was supported by the Alexander von Humboldt Foundation (Germany) for most of the completion of the project.
G.L. acknowledges support from the U.S. Department of Energy under contract number DE-AC02-76SF00515.~J.S.~is supported by NSF Grant No.~2207880.~E.V. acknowledges support from the Italian Ministero dell'Università e della Rircerca through the FIS 2 project FIS-2023-01577 (DD n. 23314 10-12-2024, CUP C53C24001460001) and through Departments of Excellence grant 2023--2027 ``Quantum Frontier'', as well as from Istituto Nazionale di Fisica Nucleare (INFN) through the Theoretical Astroparticle Physics (TAsP) project.

\bibliographystyle{bibi}
\bibliography{References}

\end{document}